\documentstyle[12pt,procsla]{article}
  
  \catcode`\@=11
  \long\def\@makefntext#1{
  \protect\noindent \hbox to 3.2pt {\hskip-.9pt  
  $^{{\ninerm\@thefnmark}}$\hfil}#1\hfill}		
  
  \def\@makefnmark{\hbox to 0pt{$^{\@thefnmark}$\hss}}  
  	
  \def\ps@myheadings{\let\@mkboth\@gobbletwo
  \def\@oddhead{\hbox{}
  \rightmark\hfil\ninerm\thepage}   
  \def\@oddfoot{}\def\@evenhead{\ninerm\thepage\hfil
  \leftmark\hbox{}}\def\@evenfoot{}
  \def\sectionmark##1{}\def\subsectionmark##1{}}
  
\begin{document}
  \rightline{CWRU-P10-96, astro-ph/9606072}
  \centerline{\normalsize\bf PROBLEMS AND CHALLENGES FOR COSMOLOGY..}
  \baselineskip=16pt
  \centerline{\normalsize\bf INVOLVING MASSIVE NEUTRINOS
  \footnote{opening lecture Venice Workshop on Neutrino Telescopes. Research
		supported in part by the DOE}}

  \centerline{\footnotesize LAWRENCE M. KRAUSS}
  \baselineskip=13pt
  \centerline{\footnotesize\it Departments of Physics and Astronomy, 
Case Western Reserve University,10900 Euclid Ave.}
  \baselineskip=12pt
  \centerline{\footnotesize\it  Cleveland OH 44106-7079  USA}
  \centerline{\footnotesize E-mail: krauss@theory1.phys.cwru.edu}
  \vspace*{0.3cm}

  \vspace*{0.9cm}
  \abstracts{I review the challenges and problems facing the standard
cosmological model, involving an $\Omega=1$ Universe dominated by non-baryonic
dark matter, which arise due to: age estimates of the universe, estimates of 
the
baryon fraction of the universe, and structure formation. Certain of these
problems are exacerbated, and certain of these are eased, by the inclusion of
some component to the energy density of matter from massive neutrinos.  I
conclude with a comparison of the two favored current cosmological models,
involving either a mixture of cold dark matter and hot dark matter, or the
inclusion of a cosmological constant}
   
  \normalsize\baselineskip=15pt
  \setcounter{footnote}{0}
  \renewcommand{\thefootnote}{\alph{footnote}}
  \section{Introduction}
  Much as it was two years ago when I was last at this meeting, the central
problem in cosmology remains how to arrange for the universe to go from its
early configuration, which looks something like this:
\begin{figure}[htb]
\includegraphics{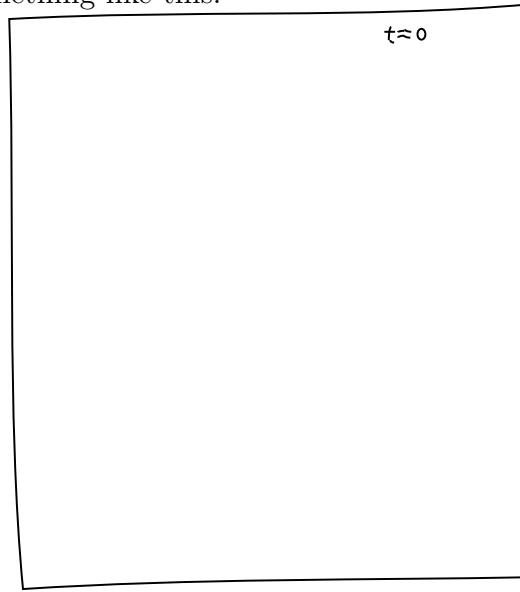}
\vglue 2.8in
\caption{The Universe near t=0}
\end{figure}

\noindent{to its present configuration, which looks something like this:}
\begin{figure}[htb]
\vglue 2.8in
\includegraphics{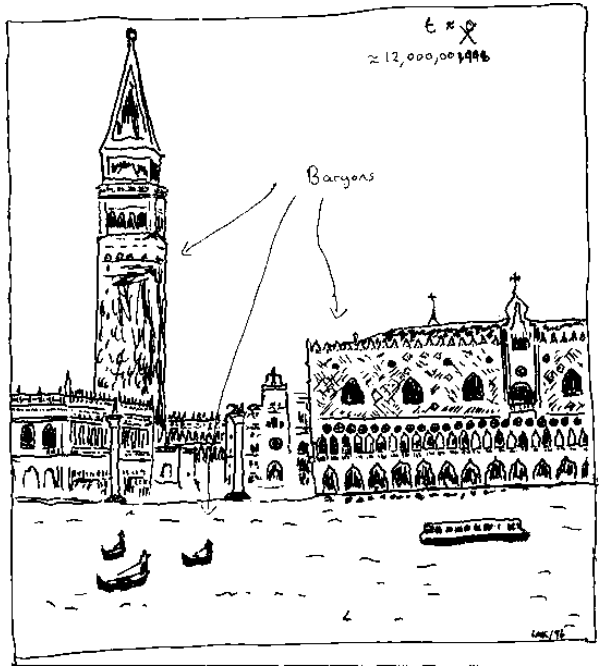}
\caption{The Universe Today}
\label{fig:largenenough}
\end{figure}

There are three aspects of the above picture which are worth examining in
detail.  First, note that there is significant structure, second that all of the
objects are baryonic, and third that the drawing captures a universe which is
at least 12 Gyr old.   All of these factors provide significant challenges to
our current understanding of cosmology, including neutrino cosmology, as I will
now describe.

  \section{Challenges for an $\Omega=1$ Universe}
  \subsection{The Age Crisis}

		As long as the Universe is decelerating (i.e. if its energy 
density
is dominated by matter or radiation) a strict upper bound on its age is given 
by:
\begin{equation}
t < {1 \over {H_0}}
\end{equation}
where $H_0 =100 h km s^{-1} Mpc^{-1}$. This result is straightforward to
understand.  If the Universe were expanding at a constant rate, then galaxies 
at
a distance d from us will have taken a time
$t=d/v ={1 \over {H_0}}$ 
to get to their current locations (where the second relation is obtained using
Hubble's Law, $v=H_0 d $).  If it is decelerating, then the time it took 
distant
galaxies to reach their present positions must be less than this time. 
This provides a strict upper limit:
$ t <12.2 Gyr [.8/h]$.

Of course, Einstein's equations allow us to solve exactly for the age in many
cosmological models.  For a flat matter dominated universe:
\begin{equation}
 t = {2 \over 3H_0} = 8.14 Gyr [.8/h]
\end{equation}
Similarly, as long as $\Omega_{matter} > 0.2$, even for an open, matter
dominated universe we have
\begin{equation}
 t < 10.4 Gyr [.8/h]
\end{equation}

The big question, therefore, is: WHAT IS $h$ ?.  With great certainty, we know
today that $0 <h <1$.  With somewhat less certainty, recent observations have
begun to narrow the range considerably.  The initial HST measurement was
\cite{freedman} : $h=.8 \pm .17$, while recent SN Ia light curve measurements
tend to yield \cite{branch,reiss} : $ h =.55-.67$.   A best estimate at the
present time might therefore be $h= .7 \pm .1 $  (recognizing that all best
estimates performed at earlier times have been different, and probably wrong,
and that this estimate may suffer the same fate..).  In this case, the above
relations imply
\vskip 0.1in 

~~~~~~~~~~~~~~~~~~~~~~~$t < 13.5 Gyr$  (open universe (= ugly)) 

~~~~~~~~~~~~~~~~~~~~~~~$t < 10.5 Gyr$ (flat universe (= beautiful))
\vskip 0.1in

The key question then becomes, Can the Universe be this Young?  To answer this,
we must resort to Cosmic Dating  (not to be confused with something which is
done in California singles bars...).   This technique is based on the following
theorem, which I will not prove here:
\vskip 0.1in

{\it Theorem: ~~  The age of the universe is greater than the age of the 
galaxy}
\vskip 0.1in

We thus merely have to determine the age of the oldest objects in our galaxy.
Among the oldest such objects are Globular Clusters, compact groups of up to
thousands of stars located throughout the galaxy.  By selecting out those
clusters which have a large halo velocity, and heavy metal abundance less than 
$0.01 \% $ of that of the sun, one is presumably sampling the among the 
oldest objects in the galaxy.

To determine the age of these objects is relatively simple, at least
in principle.  Star burn their hydrogen
fuel at a rate which goes roughly as the third power of their mass.
The time it takes therefore to burn the fuel is proportional to
the total amount of fuel ($\approx M$) divided by the rate,
or $M/M^3 \approx M^{-2}$.  Since this is a sensitive function of mass,
if one can determine which mass stars are just exhausting their fuel now
in any sample population of a fixed age, one can date the system.

Fortunately, there is a way to determine this mass.  When plotted on
a color-magnitude, or HR diagram, globular cluster stars lie in distinct
patterns, corresponding to the ``main sequence" stars, still burning
hydrogen fuel, and stars which have turned off the main sequence and
are now burning helium.   Stars of higher mass start out higher on
the main sequence, and therefore as time goes on, the ``turnoff" point
moves down the main sequence.  By fitting the observed HR diagrams with
theoretical isochrones, globular cluster ages can be determined.  
Traditionally best fits ages of 15-18 Gyr have resulted from such 
analyses.  If this is true, then there is a big age problem, even for
an open universe, if the hubble constant is larger than 55.

Since the Hubble age estimate is unambiguous for a fixed Hubble
constant the crucial uncertainty in this comparison of globular cluster and
Hubble ages resides in the
globular cluster (GC) ages estimates.  Rough arguments have
been made that changes in various input parameters in the stellar
evolution codes designed to derive globular cluster isochrones, or in
the RR Lyrae distance estimator used to determine absolute
magnitudes for GC stars, might change age estimates by 10-20 $\%$.  
However, no systematic study had been undertaken until
recently to
realistically estimate the cumulative effect of all existing
observational and theoretical uncertainties in the GC
age analysis.  This was the purpose of a year long project in which
I and my collaborators, Brian Chaboyer at CITA, Pierre Demarque at Yale,
and Peter Kernan and CWRU became involved. 

One of the such an analysis had not previously been
carried out is that it is numerically intensive.  Each
run of a stellar evolution code for a single mass point takes 
3-5 minutes on the fastest commercially available workstations.  
Nine different mass points
at three different metallicity values must be run to produce each
set of isochrones. 
If one then runs, say, 1000 different isochrone sets to explore the
different parameter ranges available, this requires over 8 weeks of
continuous processing time.

Because of the importance of this
issue, we developed the necessary Monte Carlo algorithms.  
This involved first examining the measurements of input
parameters in the stellar evolution code to determine their best fit
values, and also their uncertainties along with the appropriate
distributions to use in the Monte Carlo.  Then the stellar
evolution code and isochrone generation code were rewritten to allow
sequential input of parameters chosen from these distributions, and
output of the necessary color-magnitude (CM) diagram observables. 
Finally, we derived a fitting program to compare the predictions to
the data.   Since the numerically intensive part of this
procedure involves the Monte Carlo generation of isochrones, by
incorporating the chief observational CM uncertainty afterwards our
results can quickly be refined as this uncertainty is
refined.

We focussed on what we believe are the chief input uncertainties
in the derivation of stellar evolution isochrones.  These include: pp
and CNO chain nuclear reaction rates,
stellar opacity uncertainties, uncertainties in the treatment of
convection and diffusion, helium abundance uncertainties,
and uncertainties in the abundance of the $\alpha$-capture elements
(O, Mg, Si, S, and Ca).  Our stellar evolution code was revised to
allow batch running with sequential input of these parameters chosen
from underlying probability distributions.  We did not include the
equation of state among our Monte Carlo variables as it is now well 
understood in metal-poor main sequence stars.  

A detailed discussion of the parameter choices and methodology can
be found in our published work \cite{chab}.  
The bottom line is our result, given by 
the figure shown below.

\begin{figure}[htb]
\vglue 2.8in
\includegraphics{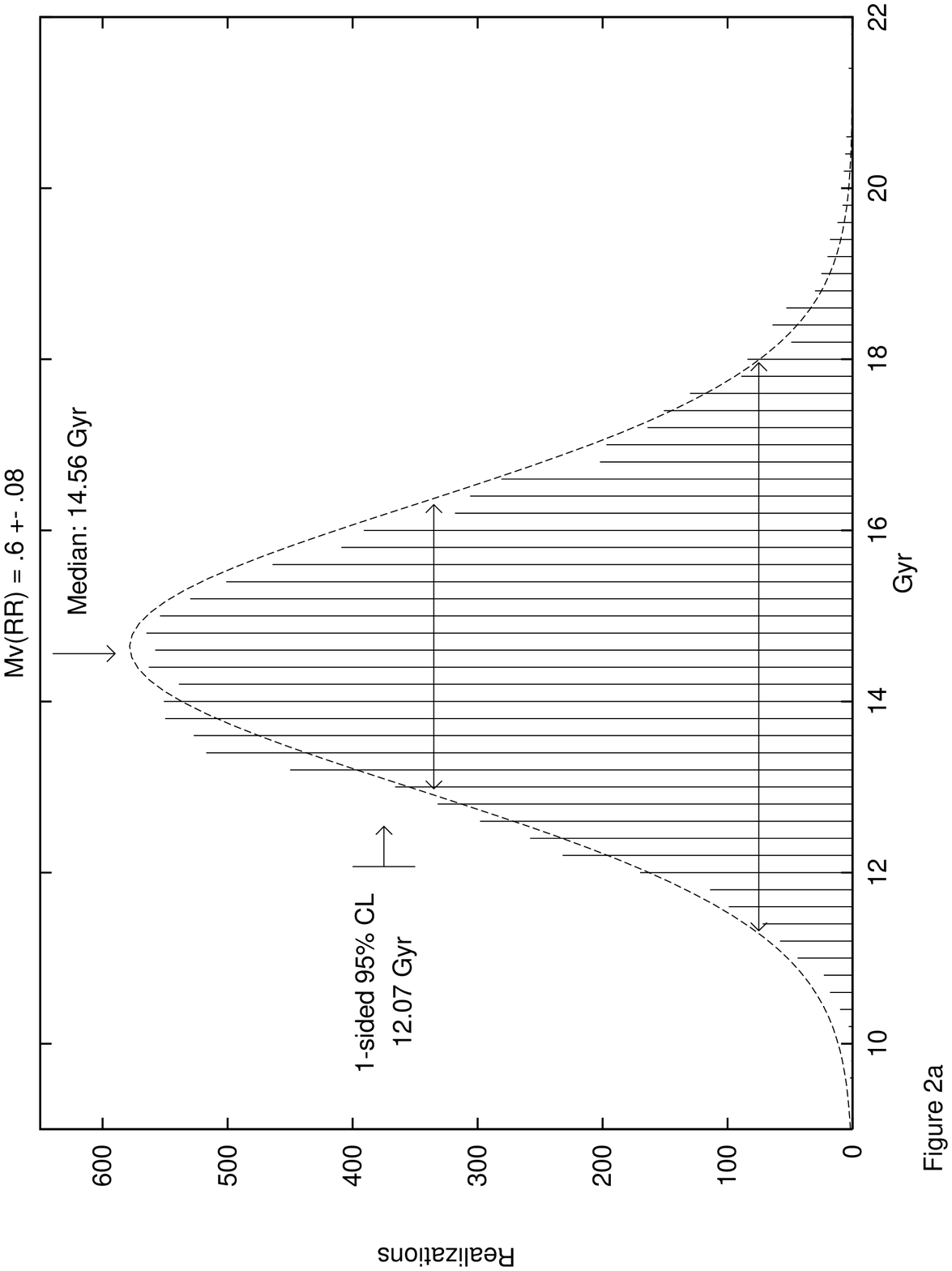}
\caption{The Distribution of Age Estimates for the Oldest Globular Clusters}
\label{fig:ageest}
\end{figure}

Our results indicate that at the one sided 95$\%$ confidence level
(determined by requiring 95$\%$ of the determined ages to fall above
this value) a lower limit of approximately 12.1 Gyr can be placed on the
mean value of these 18 Globular clusters.  (The symmetric 95$\%$
range of ages about the mean value of 14.56 Gyr is 11.6-18.1 Gyr.)  Note
that the distribution deviates somewhat from Gaussian, as one might
expect.  In particular, at the lower age limits the rise is steeper
than Gaussian, reflecting the fact that essentially all models give an
age in excess of 10 Gyr, while the tail for larger ages is larger than
gaussian.  The explicit effect of the largest single
common observational uncertainty, that in the RR Lyrae distance estimator,
 increases the net
width of the distribution by approximately $\pm 0.6$ Gyr 
(i.e. $\approx \pm 5\% $). 
Note that simply varying this over its full $2\sigma$ range,
keeping all other parameters fixed,
would produce a $ \pm 16\% $ change in GC ages estimates. 

We believe that this result can now be used with some confidence to
compare to cosmological age estimates.   Of course, in addition to the
age determined here one must add some estimate for the time it took
our galactic stellar halo to form from the initial density
perturbations present during the Big Bang expansion. Estimates for this
formation time vary from 0.1 -- 2 Gyr.  To be conservative, one can choose the
lower value.  In this case, one finds that the age of
globular clusters in our galaxy is inconsistent with a flat, matter
dominated universe unless $h < 0.54$, and for a nearly empty, matter
dominated universe unless $h < 0.80$.  If the value of h is definitely
determined to be larger than either of these values, some
modification would seem to be required. 

The most obvious modification, especially if one wants to preserve the
beauty of a flat universe, is the addition of a small, non-zero 
cosmological constant (i.e. \cite{mk}).  In this case, the relation
between Hubble constant and age of the Universe is
changed.  For $\Omega_{\Lambda} + \Omega_{matter} =1$, one finds, for example
\begin{equation}
t= (2/3) ln([1+\Omega_{\Lambda}^{1/2}]/\Omega_{matter}^{1/2}) H_0^{-1}
\Omega_{\Lambda}^{-1/2}
\end{equation}

This tends to infinity as $\Omega_{\Lambda}$ tends to unity.  Since, as
I shall argue later, a great deal of evidence suggests 
$\Omega_{\Lambda} <0.8$, this in turn implies
\begin{equation}
t \le 1.08 H_0^{-1} =13.2 Gyr (.8/h)
\end{equation}

In general, we can quantify the impact of the age problem by considering
the space of $H_0$ vs $\Omega_{matter}$ for a flat universe with a
cosmological constant.  If the age of the universe, based on the Globular 
cluster
age estimates, is $12<t<18$ Gyr, only the shaded region of phase space is
allowed  (the dashed line represents the lower bound on $H_0$ coming from 
the quote HST value with quoted error bar:

\begin{figure}[htb]
\vglue 2.8in
\includegraphics{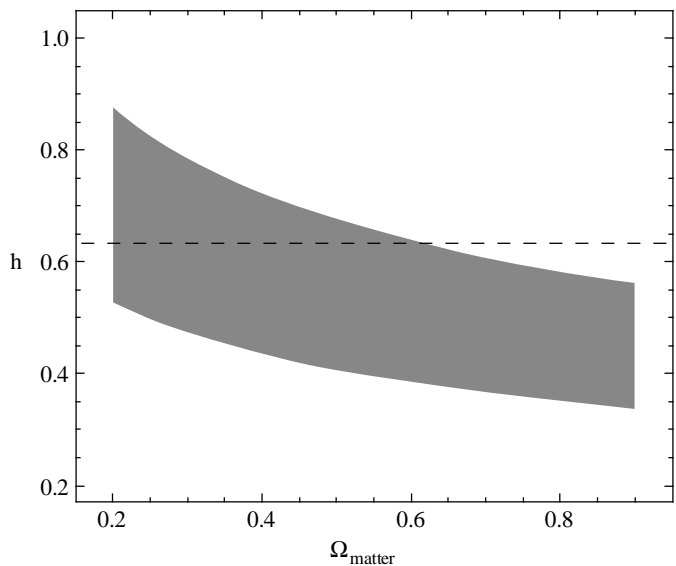}
\caption{The Constraint on $H_0$ and $\Omega_{matter}$ for a flat universe,
coming from the age of globular clusters}
\label{fig:cosmo1}
\end{figure}

\subsection{The Baryon Crisis}
  
There has been a lot of discussion in the literature recently having
to do with Crises in Cosmology...  A particularly well publicized crisis
of late is claimed to involve Big Bang Nucleosynthesis (BBN).  Now
the major crisis here may be that some people are claiming there
is a crisis.  However, in fact, BBN, makes a
well defined prediction for the upper limit on the baryon density today which 
is
not at all endangered by any new discussion of BBN uncertainties.  This
prediction be compared with other estimates of the baryon fraction of the
Universe.  It is here that there may indeed be a crisis for a flat universe,
as I shall now briefly describe.

The main virtue of BBN has not changed over the past twenty five years. 
Because the predicted abundance today of each light element produced in the
first seconds of the Big Bang explosion is a function of the abundance of
protons and neutrons in the universe, a comparison of all of the
cosmological light element abundance predictions with inferences based on
observations today can yield constraints on $\Omega_{Baryon}$.  If the allowed
range is non-zero, we call this a success of BBN.  If, on the other hand, the
allowed range is zero, we call this a crisis.

An example of the predicted range, compared to possible limits based on
inferences of ``observed" light element abundances today is shown in the
figure \cite{kk3}

\begin{figure}[htb]
\vglue 3.5in
\includegraphics{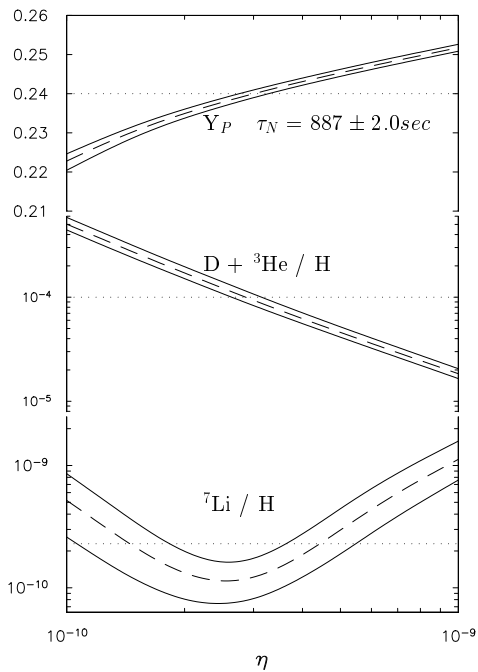}
\caption{Light element abundance predictions from BBN}
\label{fig:bbn1}
\end{figure}

Several points should be noted.  First, the predictions involve a band.  
This is
because the predicted abundances are based on model calculations which are
themselves based on measured nuclear reaction rates, which have uncertainties. 
Second, notice that the region of allowed $Y_p$, which corresponds to the mass
fraction of He, if less than the dotted line, restricts the quantity $\eta$,
which is directly related to $\Omega_{Baryon}$ to be less than some number,
while the requirement that $D+ ^3He$ is less than the dotted line restricts 
$\eta$ to be greater than some number.  Note also that these numbers are
very close... leading to the possibility of a ``crisis".

The reason there is no crisis is simply that at this point the actual
constraints coming from observations of light element abundances have huge
systematic uncertainties associated with them. (i.e. \cite{kk3,cst,ttsc}).  So,
one cannot say with any confidence that the range of allowed $\eta$ is
vanishingly small.

Given this, one might suspect that perhaps all constraints on $\eta$ coming
from BBN are suspect.  This is incorrect.  A firm upper bound on the baryon
abundance today can still be placed, because the upper limits coming from
utilizing $^4He$, $D + ^3He$, and $^6Li$  EVEN INCORPORATING maximum possible
systematic uncertainties all come together at the same value of $\eta$.  To
obviate this upper bound would require that all light element abundance
measurements are wrong, which is far more unlikely than the assumption that
some of them are wrong.....  The firm upper limit from BBN can thus be
quoted as \cite{kk3,cst}

\begin{equation}
\Omega_{Baryon} \le 0.26 h^{-2}
\end{equation}

Now, at this point you may be asking: What's this got to do with an $\Omega =1$
Universe and massive neutrinos?  The answer is that this constraint can be
compared with a lower bound on the baryon abundance today which is obtained
from measurements of X-Ray clusters of galaxies.  These objects are among the
largest structures known in the universe, and they are dominated by hot, X-Ray
emitting gas.  If one measures the temperature and luminosity of this gas as a
function of position in the cluster, and if one assumes the gas is in
hydrostatic equilibrium, and that it finds itself in a uniform, non-clumpy
potential well, then one can derive directly the depth and shape of this
potential well, and from that the total mass of the cluster.   Also, one can
derive a direct estimate of the mass of hot gas emitting the X-Ray luminosity. 
Thus, one can derive the ratio:
\vskip 0.1in
\begin{center}
$ {M_{gas} \over M_{total}} ={M_{baryon} \over M_{total}}$.
\end{center}
\vskip 0.1in

Now, IF these clusters probe the dominant mass density of the universe, 
then the above ratio
is precisely equal to $\Omega_{Baryon}$.

What makes this particularly interesting is that a number of different groups
have all recently derived this value for various rich clusters, and find that
it is rather large. (i.e. \cite{white,dwhite}).  Allowing for the quoted range,
and normalizing in the same way as I did for the BBN upper limit above, one
finds the X-Ray Cluster constraint can be expressed:

\begin{equation}
\Omega_{Baryon} \ge 0.5-.08 h^{-3/2}
\end{equation}

Comparing this constraint with the one above, one sees a potential crisis
unless either $h$ is very small, or, if X-Ray clusters are not probing all of
the mass in the universe, (or, heaven forbid, the universe is open, and not
flat---a possibility I shall not consider here in more detail because I find
it so ugly).  Indeed, if a significant fraction of the mass in the universe is
unclustered, then the X-Ray bound would only correspond to the ratio
$\Omega_{Baryon}/\Omega_{clustered matter}$.  There are two possible physical
situations which would correspond to this:

\vskip 0.1in
(1) The Cosmological Constant is Non-Zero, and the Universe is Flat:  In this
case, one derives a constraint on the $h$ vs $\Omega_{matter}$ space which is
complementary to the earlier bound, and is shown in the figure along with the
earlier bound.

\begin{figure}[htb]
\vglue 2.8in
\includegraphics{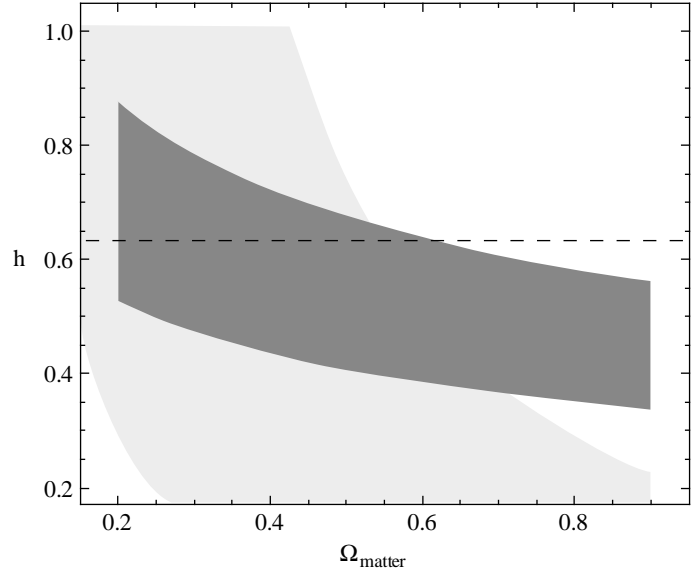}
\caption{The Constraint on $H_0$ and $\Omega_{matter}$ for a flat universe,
coming from the age of globular clusters and from baryon constraints}
\label{fig:cosmo2}
\end{figure}

\vskip 0.1in
(2) There is unclustered matter (i.e. neutrinos!!!):  Finally, for the first
time in this lecture, I have referred to neutrinos directly.  Light neutrinos,
because they are ``hot", i.e. relativistic, until relatively late times, are
not efficiently captured in clusters.  Depending upon the mass of the
neutrinos, and the fact that current mixed dark matter models (see following
discussion) suggest a small admixture of light neutrinos along with cold dark
matter in a flat universe, several authors have argued that this might
alleviate the baryon crisis  \cite{kkpn,ss}.  However, both groups have found
that quantitatively, things are only marginally improved.  By allowing most of
the light neutrinos to escape from galaxies, one finds that the X-Ray estimates
for $\Omega_{Baryon}$ might overestimate the actual value by up to $25 \%$. 
For acceptably small values of $h \approx 0.5$, this could resolve the crisis.

  \section{Challenges for a Flat Universe Involving Massive Neutrinos: Large
Scale Structure}

Now that I have finally introduced cosmological neutrinos as dark matter, I can
introduce the third major crisis in modern cosmology which has revised the
accepted orthodoxy regarding the prejudice of theorists of the nature of dark
matter.  This crisis has to do with Large Scale Structure. Put succinctly,
the historical evolution of the theoretical Best fit model of the universe has
been as follows:

\vskip 0.2in

\noindent{ 1981:~~~~~~~~  $\Omega =1;  \Omega_{\nu} =1 ;  h \approx 0.5$

\vskip 0.2in

\noindent{ 1985:~~~~~~~~  $\Omega =1;  \Omega_{CDM} =1 ;  h \approx 0.5$

\vskip 0.2in

\noindent{ 1995:~~~~~~~~  $\Omega =?;  \Omega_{CDM} <1 ;  h > 0.5$

\vskip 0.2in
As can be seen, current wisdom now allows, indeed perhaps requires, something
other than Cold Dark Matter and Baryons in the Universe.  Whether this
something extra is neutrinos, or a cosmological constant, will perhaps be
determined by further measurements of large scale structure.

Now the subject of Large Scale Structure and primordial density perturbations 
is
far too complex to treat with any justice here.  However, there is one aspect
which gives an important constraint and which is relatively model independent
while also being easily explained.  

If one is going to posit, a priori, a spectrum of primordial density
fluctuations, then the fourier space estimate would be of the following form:

\begin{equation}
({{\delta \rho} \over \rho})^2 \approx k^n
\end{equation}

The reason for this is simple.  Anything but a power law would pick out some
preferred scale at early times, and it seems unreasonable to expect that a
scale of cosmologically interesting size would be fixed by the microphysics
near t=0.  Now, even before Inflationary model predictions, it was suggested
that $n \approx 1$.  This is because if n deviated significantly from unity,
one would predict either too many primordial small black holes, or too large an
anisotropy on large scales today.  

Now, the above picture is that of the spectrum of primordial density
fluctuations.   However, the spectrum of fluctuations which eventually leads to
galaxy formation is not the primordial one, but one which has rather been
evolved by causal physics.  Causality provides an important bit of structure. 
In particular, as long as there is any radiation in the universe today, the
energy density of the Universe was once, at earlier times, dominated by this
radiation.  Thus, if the universe is matter dominated today, for all times
earlier than some time $t_{eq}$ it was radiation dominated.  Associated with
this time will be some fourier mode $k_{eq}$, the wavelength of which is equal
to the size of the horizon at this time.  All modes with larger wavenumber
will have a wavelength smaller than the horizon size, and can thus have been
affected by earlier causal microphysical processes.  It turns out that during a
radiation dominated expansion, density fluctuations do not grow, and in fact
can often be damped.  

The net result of this causal behavior is that an initial spectrum which
grows monitonically with $k$, can instead be transformed into a spectrum
which is reduced somewhat starting around $k_{eq}$---hence there is some
curvature to the evolved spectrum of density fluctuations for wavelengths
corresponding to the region of $k_{eq}$.  By measuring the two point
correlation function of galaxies, one can hope to explore the evolved
spectrum in this region, and compare it to predictions.  The precise value
of $k_{eq}$ will depend upon the time of matter-radiation equality, and hence
upon the matter density today.   Specifically, since $\rho_{matter} \approx
\Omega_{matter} h^2$, and the relation between fourier modes and physical
wavelengths scales as $1/h$, one finds that the value of $k_{eq} $ depends upon
the combination $ \Omega h$.     Many different measurements of the shape of
the spectrum of density fluctuations in the universe yield the constraint:

\begin{equation}
0.2 < \Omega_{matter} h < 0.3 
\end{equation}

This is perhaps the single most significant constraint coming from observations
of Large Scale Structure which has altered our view of what might be the
favored cosmological model in the past decade.  Several options come to
mind:

\vskip 0.2in

(1) $\Omega_{matter} < 1$, and purely CDM.  This implies either an ugly (open)
universe, or a beautiful, flat universe, with a cosmological constant.  Adding
the constraint from the above equation to our earlier constaints then gives the
following picture (where I have also introduced the constraint
$\Omega_{matter} \ge 0.3$, as suggested by analyses of large scale
peculiar velocity fields.)  All of these constraints give a consistent
picture\cite{mk} for $ 0.5 < \Omega_{\Lambda} < 0.7$.  

\begin{figure}[htb]
\vglue 2.8in
\includegraphics{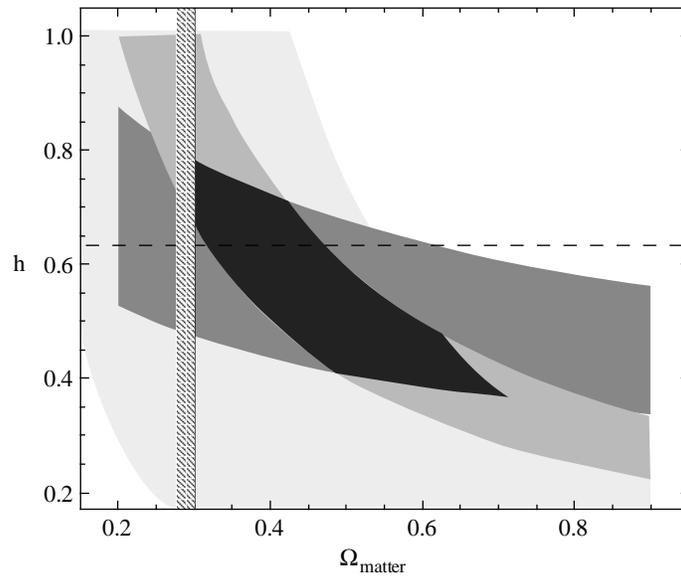}
\caption{The Constraint on $H_0$ and $\Omega_{matter}$ for a flat universe,
coming from the age of globular clusters, from baryon constraints, and from the
power spectrum of density fluctuations at galaxy scales and larger, assuming an
n=1 spectrum of primordial density fluctuations and no significant component of
hot dark matter }
\label{fig:cosmo3}
\end{figure}

(2)  Change the shape of the primordial spectrum so that the shape near
$k_{eq}$ shifts.  This is possible if $ n \ne 1$, for example.
\vskip 0.1in

 OR, 
\vskip 0.1in

(3)  Change the value of $k_{eq}$ by changing the time of matter radiation
equality:  Both examples of such a behavior involve massive neutrinos.  
Prof. Sciama, who is was at this meeting, 
has proposed a scenario involving decaying neutrinos. 
Alternatively, by adding a hot component to the dark matter density, as one
would get from light stable neutrinos, one will also change the epoch of matter
radiation equality.

  \section{CHDM vs $\Lambda$CDM}
  
If one is to preserve a flat universe today, it seems clear that one is being
driven therefore by existing cosmological observations 
to one of two extremes.  Either the
Universe is dominated by a Cosmological Constant, or some fraction of the dark
matter dominating the mass density of the universe today is in the form of
light neutrinos.  These two models present what have become the favored
scenarios at the turn of the millenium.  Whether either survives into the next
millenium will depend upon how successfully various existing challenges are
overcome.

\vskip 0.2in

\noindent{(1) {\bf LSS Challenges for CHDM}}

\vskip 0.2in
(a)  Early Galaxy formation

(b) The detailed Shape of the Power Spectrum

(c) the Void Probability function

(d) The Damped Lyman- $\alpha$ Forest.

\vskip 0.2in
All of these challenges come down to the same issue.  Hot dark matter
suppresses the growth of fluctuations on smaller scales.  This will have the
effect of causing galaxies to form later, reducing the magnitude of the power
spectrum at small scales, changing the probabilities of large voids, and
reducing the likelihood of producing significant clumped hydrogen clouds.  Many
authors have recently analyzed these issues, and at least 
two sets\cite{primack,ma} have suggested that these
challenges can be successfully met, if the density of HDM is approx
$\Omega_{\nu} \approx 0.2$, and if there are two species of light neutrinos
both with a mass near 2 eV.
\vskip 0.2in

\noindent{(2) {\bf LSS Challenges for $\Lambda$CDM}}

\vskip 0.2in
(a)  Non-Linear Power on Small Scales

(b) The Deceleration Parameter
  \vskip 0.2in

A cosmological constant dominated universe avoids the problem of the growth of
small scale structure associated with mixed dark matter models.  However, by
avoiding this problem too efficiently, it might result in excessive structure 
on
small scales.  Recent numerical models have suggested, at least for the
extreme value of $\Omega_{\Lambda} \approx 0.8$, that unacceptably large
galaxy potential wells will form.  Perhaps the biggest observational challenge
which may arise in the near term for a cosmological constant dominated
universe is the fact that such a universe will not be decelerating.  As a
result, if a careful measurement of the deceleration parameter becomes
possible, as recent observations of SN 1a light curves suggest might be the
case, one could rule out this scenario altogether.  Indeed, very preliminary
observations of Type 1a SN do suggest that the deceleration parameter is
positive, which would require that the cosmological constant contribution to
the energy density today not exceed that due to matter.

  \section{Conclusions}
  Which of these favored scenarios, if either, is more appealing?  Well, beauty
is in the eye of the beholder.  To help guide the eye, however, I present my
own scorecard, using the consumer's digest convention, where an open circle is
very bad, and a closed circle is very good.  In this way, I compare
CHDM,$\Lambda$CDM, and CDM flat models against constraints from age, the
baryon abundance, the shape of the power spectrum of density fluctuations,
small scale structure, and whether the model is theoretically contrived. 

\begin{figure}[htb]
\vglue 4.7in
\includegraphics{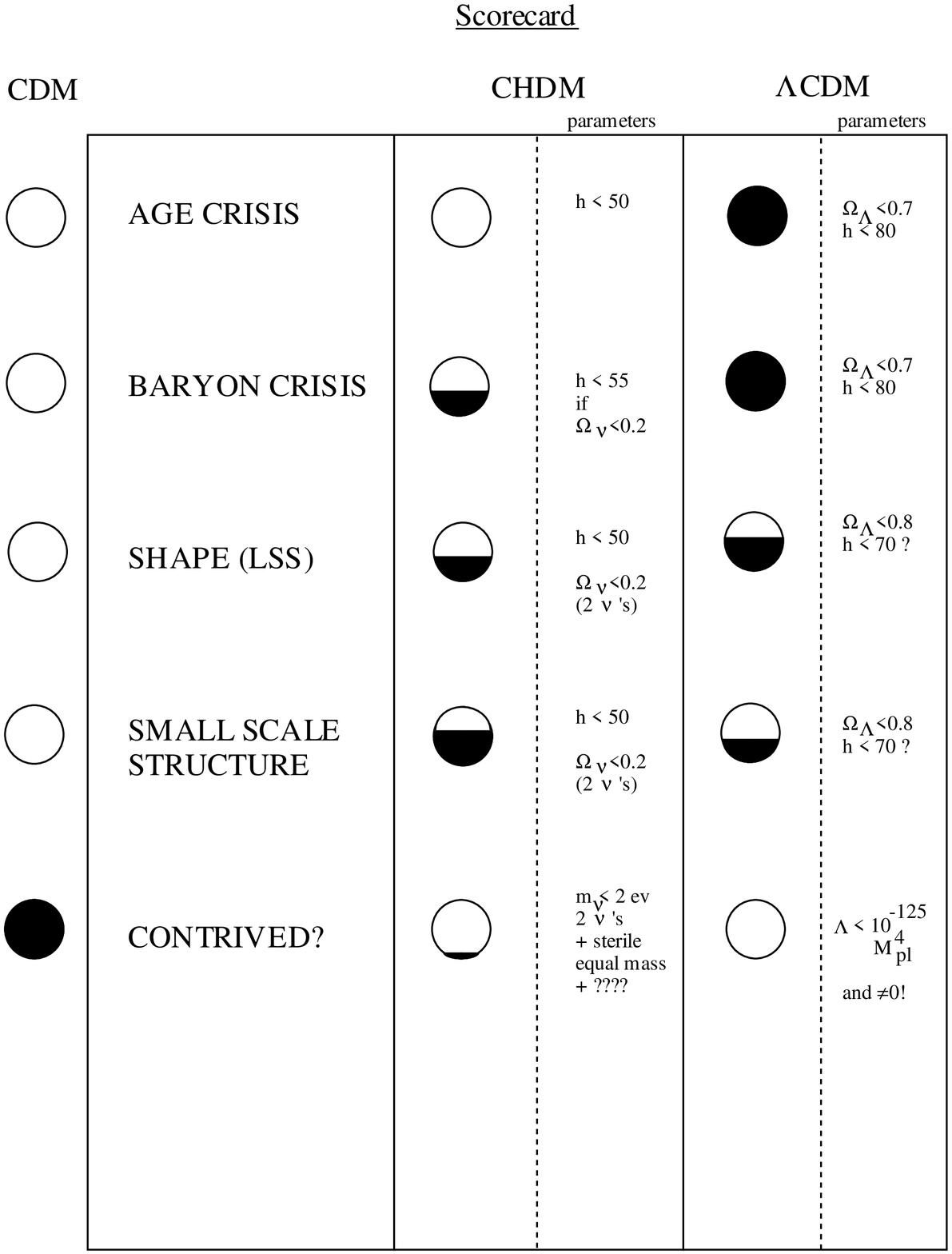}
\caption{A Consumer's Guide to Cosmological Models}
\label{fig:consum}
\end{figure}

I have no idea whether any of these models will survive the test of time,
and observations.  However, at the moment the choice seems clear: 
{\it Neutrinos, or Nothing!}

\section{Acknowledgments}
I would like to thank the organizers of the meeting for their kind
hospitality and patience.


\section{References} 
  \begin{thebibliography}{9}
  \bibitem{freedman} W. Freedman {\it et al}, {\it Nature} 
{\bf 371} (1994) 757.
  \bibitem{branch}  D. Branch {\it et al}, {\it astro-ph/9604006}
preprint 
  \bibitem{reiss} A.G. Reiss {\it et al}, {\it Ap. J.} {\bf 438} (1995) L17
  \bibitem{chab} B. Chaboyer, P. Demarque, P. J. Kernan, L. M. Krauss, 
{\it Science}, {\bf 271} (1996) 957
  \bibitem{mk} L. M. Krauss and M. S. Turner, {\it J. Gen. Rel. Grav.} 
{\bf 27} (1995) 1137
  \bibitem{kk3} L. M. Krauss and P.J. Kernan, {\it Phys. Lett.} 
{\bf B347}, (1995) 347
  \bibitem{cst} C. Copi, D.N. Schramm, and M. Turner, {\it Science}
{\bf 267} (1995) 192
  \bibitem{ttsc} M. Turner {\it et al}, Fermilab preprint 1996
  \bibitem{white} S. D. M. White {\it et al}, {\it Nature}
{\bf 366} (1993) 429
  \bibitem{dwhite} D. A. White and A. C. Fabian, {\it M.N.R.A.S.}
{\bf 273} (1995) 73
  \bibitem{kkpn}  L. Kofman {\it et al} ,{\it UH-IfA -95/46} preprint
  \bibitem{ss} R. W. Strickland and D. N. Schramm, {\it Fermilab pub}
 95/398-A
  \bibitem{primack}  R. Somerville, J. R. Primack, R. Nolthenius, 
preprint, {\it astro-ph/9604051} 
  \bibitem{ma} C-P Ma, {\it Ap. J}, to appear Nov 1996
  \end{thebibliography}
  \end{document}